\documentstyle[aps,epsfig,floats,preprint]{revtex}
\newcommand{\labx}[1] 
{
\label{#1}}

\begin{document}
\draft
\tighten

\def\lsi{\raise0.3ex\hbox{$<$\kern-0.75em\raise-1.1ex\hbox{$\sim$}}}
\def\gsi{\raise0.3ex\hbox{$>$\kern-0.75em\raise-1.1ex\hbox{$\sim$}}}
\newcommand{\lsim}{\mathop{\lsi}}
\newcommand{\gsim}{\mathop{\gsi}}
\newcommand{\x}{\vec{x}}
\newcommand{\y}{\vec{y}}
\newcommand{\k}{\vec{k}}
\newcommand{\B}{\vec{B}}
\newcommand{\hxi}{\hat{\xi}}
\newcommand{\hk}{\hat{k}}
\newcommand{\hx}{\hat{x}}
\newcommand{\hi}{\hat{\imath}}
\newcommand{\hj}{\hat{\jmath}}

\title{Phase transition dynamics in the hot Abelian Higgs model}

\author{M.~Hindmarsh$^1$ and A.~Rajantie$^2$}
\address{
$^1$Centre for Theoretical Physics,
University of Sussex,
Brighton BN1 9QJ, 
United Kingdom
\\
$^2$DAMTP, CMS, Wilberforce Road, Cambridge CB3 0WA, United Kingdom
}

\date{March 29, 2001}
\maketitle

\begin{abstract}
We present a detailed numerical study of the
equilibrium and 
non-equilibrium dynamics of the phase transition in the
finite-temperature Abelian
Higgs model.
Our simulations use classical equations of motion both with
and without hard-thermal-loop corrections, which take into account the
leading quantum effects. 
From the equilibrium real-time
correlators, we determine the Landau damping rate, the plasmon
frequency and the plasmon damping rate. 
We also find that, close to the phase transition, 
the static magnetic field correlator shows
power-law magnetic screening at long distances. 
The information about the damping rates allows us to
derive a quantitative prediction for the number density of topological
defects formed in a phase transition. We test this prediction in a
non-equilibrium simulation and show that the relevant time scale for 
defect formation is given by the Landau damping rate.\\
\\
DAMTP-2001-28 SUSX-TH-01-015
\end{abstract}

\section{Introduction}
Whilst there are many useful techniques for studying the
equilibrium properties of finite-temperature field theories, understanding the
non-equilibrium dynamics is a much harder task. 
Nevertheless, it would be essential for many fields of physics, 
for instance
cosmology, heavy ion physics and condensed matter
physics. In all these fields, new empirical
data will be
available in near future, which would allow the theories to be tested,
but the complexity and the non-equilibrium nature of the phenomena
make it difficult to derive theoretical predictions that could be
compared with the data.

One fairly generic consequence of phase transitions is formation
of topological defects~\cite{Kibble:1976sj,Hindmarsh:1995re}. 
If the phase transition is associated with a
spontaneous breakdown of a global symmetry, this process is well
understood. The correlation length of the order parameter cannot
keep up with its equilibrium value, which diverges at the transition
point. The direction of the symmetry breaking must therefore 
be uncorrelated at
long distances, and at places where these correlated domains meet,
topological defects are formed. This is called the
Kibble-Zurek (KZ) mechanism (see e.g.~\cite{Zurek:1996sj} for a review).

If the symmetry that gets broken is a local gauge invariance, the
above argument cannot be used directly, because the direction of the
order parameter is not a gauge invariant quantity. We studied this
recently in the context of the Abelian Higgs
model~\cite{Hindmarsh:2000kd}, and
pointed out that the thermal fluctuations of the magnetic field lead
to another mechanism that forms topological defects.
The argument was based on fairly generic assumptions, but leads to
some concrete predictions that were confirmed in numerical
simulations.

The aim of this paper is to study in more detail the dynamics of the
Abelian Higgs model during the phase transition from the Coulomb phase
to the Higgs phase. In particular, we concentrate on those degrees of
freedom that are relevant for defect formation. This allows us to test
the scenario of Ref.~\cite{Hindmarsh:2000kd} on a more quantitative level.

The theory considered in Ref.~\cite{Hindmarsh:2000kd} was classical, 
and although
the same arguments apply to the quantum theory as well, the details of
the dynamics are different. The full quantum field theory
cannot be simulated in practice, but one can argue that the dynamics of
the relevant long-wavelength degrees of freedom are  
classical~\cite{Grigoriev:1988bd}. By
integrating out the short-wavelength fluctuations perturbatively, one
obtains a classical effective theory with non-local
interactions~\cite{Pisarski:1989vd,Braaten:1992gm,Blaizot:1993zk%
,Moore:1998sn}, 
to which we refer as the hard-thermal-loop (HTL)
improved theory. In order to understand how the quantum effects change
the dynamics, we simulate this HTL improved theory using the
method developed in Ref.~\cite{Rajantie:1999mp}.

The structure of the paper is the following: in Sect.~\ref{sect:ahm} we 
present both the classical and HTL improved Abelian Higgs models.  In 
Sect.~\ref{sect:defect} we discuss defect formation in the model, 
comparing the mechanism presented in \cite{Hindmarsh:2000kd} with the 
Kibble-Zurek scenario.  In Sects. \ref{sect:classsim} and \ref{sect:htlsim} we
describe our numerical simulations and present the results.  Conclusions are
given in Sect.~\ref{sect:conc} and technical details of the HTL improved
equations of motion and the lattice formulation in the two Appendices.

\section{Abelian Higgs model}
\label{sect:ahm}
The Abelian Higgs model is defined by the Lagrangian
\begin{equation}
{\cal L}=-\frac{1}{4}F_{\mu\nu}F^{\mu\nu}+|D_\mu\phi|^2
-m^2|\phi|^2-\lambda|\phi|^4,
\labx{equ:lagr}
\end{equation}
where $D_\mu=\partial_\mu+ieA_\mu$ and 
$F_{\mu\nu}=\partial_\mu A_\nu-\partial_\nu A_\mu$.

A particularly interesting feature of this theory is the existence of
Nielsen-Olesen vortex solutions \cite{Nielsen:1973cs}. These string-like 
topological defects are characterized by a zero of the Higgs field
at the centre of the vortex around which the Higgs phase angle
has a non-zero winding number
\begin{equation}
n_C=\int_Cd\vec{r}\cdot\nabla\gamma(\vec{r})\neq 0.
\end{equation}
Here $C$ is a closed path around the vortex 
and $\gamma$ is the Higgs phase angle,
i.e.,~$\phi=|\phi|\exp(i\gamma)$.

At finite temperature, perturbation theory is plagued by
infrared divergences 
\cite{Linde:1980ts},
which can be partly cured by a resummation of the perturbative expansion,
but even the resummed expansion breaks down near the 
transition.
Static equilibrium
quantities, such as the phase diagram of the theory, 
can still be calculated reliably with non-perturbative
Monte Carlo simulations.

The model has a phase transition between the high-temperature
Coulomb phase and the low-temperature Higgs phase
at $T^2=T_c^2\approx 12(-m^2)/(3e^2+4\lambda)$.
In the perturbative regime ($\lambda\ll e^2$), 
the transition is of first order~\cite{Dimopoulos:1998cz}, 
and at larger $\lambda$ it becomes 
continuous~\cite{Kajantie:1998hn}.
There are no local order parameters, but a number
of non-local ones: the photon mass and the vortex tension are non-zero
in the broken phase and vanish at the transition~\cite{Kajantie:1999zn}. 
Therefore the
transition is not a smooth crossover like the electroweak phase transition
\cite{Kajantie:1997qd}.

Monte Carlo simulations cannot be used for real-time quantities in the
quantum theory, because
the necessary path integral is not Euclidean but consists of a 
complicated path in complex time \cite{Niemi:1984nf}. 
However, we can utilize the fact that modes with different momenta
behave in very different ways \cite{Grigoriev:1988bd}. 
The soft, long-wavelength modes ($k\ll T$)
have large occupation numbers, and they can be approximated very well
by a classical theory.
This makes numerical simulations feasible, because the time-evolution
of a classical field theory
can be found simply by solving the equations of motion numerically.

\subsection{Classical theory at finite temperature}

Classical field theory at finite temperature is
ultraviolet divergent, and thus the results depend on the lattice
spacing $\delta x$. 
Divergences like these are generic to all low-energy
effective theories, and are exactly cancelled by corrections the
high-momentum modes induce to the effective Lagrangian.
If one is only interested in static equilibrium quantities, these
corrections can be calculated in the limit of 
high temperature and small lattice
spacing $\delta x$~\cite{Laine:1995ag,Laine:1998dy},
\begin{equation}
\labx{equ:mTren}
m_T^2= m^2+(3e^2+4\lambda)\left(
\frac{T^2}{12}-
\frac{3.176 T}{4\pi\delta x}\right).
\end{equation}
The approach we use in Sect.~\ref{sect:classsim} is to take this correction into
account and solve the classical equations of motion
\begin{eqnarray}
\partial_\mu F^{\mu\nu}&=&
-2e{\rm Im}
\phi^*D^\nu\phi,
\nonumber
\\
D_\mu D^\mu\phi&=&-m_T^2\phi-2\lambda(\phi^*\phi)\phi.
\labx{equ:eom}
\end{eqnarray}
However, as was pointed out in Ref.~\cite{Bodeker:1995pp}, 
these equations do not reproduce the real-time dynamics
of the quantum theory correctly. 
Thus, the results are not reliable on 
a quantitative level, but they can still give a reasonably good
qualitative picture of the dynamics, and provide a 
non-trivial test for the scenario presented in
Section~\ref{sect:defect}.

\subsection{HTL improved theory}
\label{sect:htltheory}
If the couplings are small, the system is close to thermal equilibrium
and $T\delta x\gg 1$, it is
possible to construct a classical theory which
approximates the dynamics of the original quantum theory to
leading-order accuracy in the coupling constants~\cite{Moore:1998sn}.

Near the phase transition,
$m^2\sim -e^2T^2$, and we can use the high temperature
approximation $T\gg m$ in our loop integrals, provided $e$ is small.
We calculate the one-loop
corrections from the hard modes to the self-energies of $\phi$ and $A_i$,
and resum them into the effective 
Lagrangian \cite{Kraemmer:1995az}
\begin{eqnarray}
{\cal L}_{\text{HTL}}&=&-\frac{1}{4}F_{\mu\nu}F^{\mu\nu}
-\frac{1}{4}m_D^2
\int\frac{d\Omega}{4\pi}F^{\mu\alpha}
\frac{v_\alpha v^\beta}{(v\cdot\partial)^2} F_{\mu\beta}\nonumber\\
&&+|D_\mu\phi|^2
-m_T^2|\phi|^2-\lambda|\phi|^4,
\labx{equ:lagrHTL}
\end{eqnarray}
where $m_T^2$ is given by Eq.~(\ref{equ:mTren}), and the integration 
is taken over the unit
sphere of velocities $v=(1,\vec{v})$, $\vec{v}^2=1$.
The Debye mass has the value $m_D^2=\frac{1}{3}e^2T^2+\delta m_D^2$,
where 
$\delta m_D^2$ is a counterterm that cancels the UV divergences
and is discussed in more detail in Appendix~\ref{app:latthtl}.

All the degrees of freedom remaining in this effective theory are classical,
and therefore it can be treated as a classical theory.
The time evolution of the fields will then be determined by
the equations of motion [cf.~Eq.~(\ref{equ:eom})]:
\begin{eqnarray}
\partial_\mu F^{\mu\nu}&=&m_D^2\int\frac{d\Omega}{4\pi}
\frac{v^\nu v^i}{v\cdot\partial}E^i-2e{\rm Im}
\phi^*D^\nu\phi,
\nonumber
\\
D_\mu D^\mu\phi&=&-m_T^2\phi-2\lambda(\phi^*\phi)\phi.
\labx{equ:nonloc}
\end{eqnarray}
As such, the equations of motion (\ref{equ:nonloc}) are 
not well suited for our purposes. Firstly, 
it is not obvious how to find the corresponding Hamiltonian,
which is necessary for preparing the initial configurations. And secondly,
the equations of motion are non-local both in space and
time.

In Ref.~\cite{Rajantie:1999mp}, a convenient local formulation was 
presented along
the lines of Ref.~\cite{Blaizot:1993zk}. The latter work 
introduces a new local field
$W(t,\vec{x},\vec{v})$, representing the departure from the equilibrium
distribution function for hard particles with velocity $\vec{v}$. 
Since the hard particles move at the speed of light, $\vec{v}$ consists of
two free coordinates, making $W$ a six-dimensional field.
In our formulation, part of the velocity dependence decouples from the
dynamics of the soft modes, and we can describe exactly the same
dynamics
with two five-dimensional fields, 
$\vec{f}(t,\vec{x},z)$ and
$\theta(t,\vec{x},z)$, where $z\in[0,1]$ is 
the cosine of the angle
between the hard mode velocity and
the gradient of the distribution function.
The details of this formulation are given in Appendix~\ref{app:htleom}.

In order to simulate the model numerically, we define
the theory on a periodic spatial lattice, and 
the $z$ dependence of $\vec{f}$ and $\theta$ is  discretized by 
expressing them as sums over a finite number of Legendre
modes, whose order ranges from $0$ to $N_{\rm max}$ \cite{Rajantie:1999mp}. 
The resulting equations of motion are detailed in Appendix \ref{app:htleom}. 
We merely note here that, as shown in Ref.~\cite{Rajantie:1999mp}, 
for a
given value of $N_{\rm max}$ and a given momentum $k$, 
the approximation breaks down at times 
\begin{equation}
t\gsim t_c(k) = 4N_{\rm max}/k.
\labx{equ:crittime}
\end{equation}
Therefore one can strictly speaking only measure correlators up to
$\Delta t\sim N_{\rm max}\delta x$, but since the modes that are in
equilibrium will remain in equilibrium even beyond that, one can
essentially trust the results until $\Delta t\approx t_c(k)$, where
$k$ is the relevant momentum scale. 

In Sect.~\ref{sect:htlsim}, we study the dynamics of the model 
using the HTL improved
equations of motion, and describe how these corrections
change the results from the classical case.

\section{Defect formation}
\labx{sect:defect}
In cosmology, temperature of the universe decreases as a result of its
expansion, and this leads to phase transitions. If we write the equations of
motion in conformal coordinates, the effect of the expansion
can be absorbed completely
into a varying zero temperature Higgs mass term, $m^2=m^2(t)$. 

In order to see this, we perform the conformal rescaling $dx^\mu \to adx^\mu$,
$v^\mu \to av^\mu$, $A_\mu \to a^{-1}A_\mu$ and $\phi \to a^{-1}\phi$, where
$a=a(t)$ is the scale factor of the Universe.  If we can neglect the expansion
rate in comparison to the microscopic timescales of the theory, which are given
by $m_T^{-1}$ and $m_D^{-1}$, the effect of this rescaling on the action is
to replace the
mass terms $m_T^2$ and $m_D^2$ by $(m_Ta)^2$ and $(m_Da)^2$ 
respectively.   The hard modes are assumed to be ultra-relativistic, 
so they will stay close to a thermal distribution
with temperature $T = \bar{T} a^{-1}$.  Hence $(m_Da)^2$ stays constant, as do
the rescaled thermal corrections to the Higgs mass, leaving the only
time dependence in the parameter $m^2(t) = m^2a^2(t)$.
In the following, we
will assume that the phase transition is triggered in this way, 
by a mass term
decreasing below a critical value, but we believe that the qualitative
features would not be very different if some of the other parameters
were changing at the same time.

When the system enters the Higgs phase, Nielsen-Olesen vortices are
formed. In the limit $e\rightarrow 0$, this can be understood in terms
of the Kibble mechanism~\cite{Kibble:1976sj}. 
When $m^2$ approaches its critical value, the Higgs
correlation length $\xi$ grows, and if the system could remain in
equilibrium it would eventually diverge at the transition
point. However, $\xi$ cannot grow arbitrarily fast, because at the
very least it is constrained by the finite speed of light.
Therefore it remains finite if
the transition takes place in finite time. At the transition point,
the system consists of correlated domains of size $\hat{\xi}$
determined by the maximum correlation length reached. In each of these
domains, the phase angle of the Higgs field is chosen 
independently of all others, and this gives rise to frustrations,
vortices, where these domains meet. Up to a numerical factor, the
number of vortices piercing a unit area is
\begin{equation}
n=N/A\approx \hat{\xi}^{-2}.
\labx{equ:zurekdensity}
\end{equation}

In practice, the maximal rate of change of the correlation length may
be well below the speed of light, and this argument can indeed be made
more precise by considering the dynamics of the system in more
detail~\cite{Zurek:1993ek,Zurek:1996sj}.

In Ref.~\cite{Hindmarsh:2000kd}, 
we argued that this picture is not complete if $e>0$.
After all, the phase angle of the Higgs field is not gauge
invariant, and therefore arguments based on it cannot apply. 
Nevertheless, if the amplitude of the
magnetic field is small, we can fix the gauge in which
$A_i\approx0$, and then we can use the above picture in this gauge. In
these cases the above Kibble-Zurek
scenario should work. However, if the initial state is at a non-zero
temperature, magnetic field is never exactly zero, and it must be
taken into account.

In the symmetric phase, the thermodynamics of the gauge field is
described by ordinary electrodynamics. The energy of any magnetic field
configuration is approximately given by
\begin{equation}
H_{\rm EM}[\vec{B}(\vec{x})]=\frac{1}{2}\int d^3x\vec{B}(\vec{x})^2,
\end{equation}
and the probability with which the thermal fluctuations can generate
that configuration is proportional to $\exp(-H_{EM}/T)$.
Let us consider a circular loop $C$ that bounds a surface $S$. 
The magnetic flux
through the surface is
\begin{equation}
\Phi_S=\int d{\vec S}\cdot \vec{B}.
\end{equation}
Although this is zero on the average, it is non-zero in almost all
configurations, and therefore the typical value, given by 
$\sqrt{\Phi_S^2}$ is non-zero. We can estimate its value by
calculating the energy of the field configuration that minimizes the
energy for a given value of $\Phi_S$. This minimal configuration is
simply a magnetic dipole, and its energy is
\begin{equation}
E_{\rm min}(\Phi_S)\approx \Phi_S^2/R,
\end{equation}
where $R$ is the radius of the loop $C$.
Thermal fluctuations can create this configuration if 
$T\gsim E_{\rm min}(\Phi_S)$, and solving this for $\Phi_S$ shows that
the typical flux through the loop is
\begin{equation}
\Phi_S\approx \sqrt{TR}.
\end{equation}

When the system enters the Higgs phase, magnetic flux is confined into
flux tubes, which costs energy. Therefore the dynamics tries to
decrease the magnetic flux, but it can only do so at short
distances, because the range of the interaction between the flux tubes
decreases rapidly.

We can be more specific in Fourier space. 
We can write the two-point correlator of the magnetic flux density
$\vec{B}$ as
\begin{equation}
\langle B_i(\vec{k})B_j(\vec{k}')\rangle
=\left(\delta_{ij}-\frac{k_ik_j}{k^2}\right)(2\pi)^3
\delta(\vec{k}+\vec{k}')G(k).
\labx{equ:spatcorr}
\end{equation}
In the symmetric phase,
different
Fourier modes $\vec{B}(\vec{k})$ behave as independent oscillators in thermal
bath, and thus each of them has the same amplitude $G_0(k)=T$.
When the system enters the broken phase, magnetic field becomes
massive, and the equilibrium distribution changes into
\begin{equation}
G(k)=T\frac{k^2}{k^2+m_\gamma^2}.
\labx{equ:scpert}
\end{equation}
In order to stay in equilibrium, the amplitude of the long-wavelength
modes must drop rapidly, but the time scale $\tau(k)$ 
of the dynamics of these
modes is very slow. It depends on the individual system, but in
general
$\lim_{k\rightarrow 0}\tau(k)=\infty.$ Thus the modes with $k$ less
than some critical value $\hk$ cannot remain in equilibrium.
If we know $\tau(k)$, we can calculate $\hk$ from the condition
\begin{equation}
\left|\frac{d\ln G(\hk)}{dt}\right|\approx\frac{1}{\tau(\hk)}.
\labx{equ:adiabatic}
\end{equation}

The consequence of the above process is that
there will be long-wavelength magnetic fields present even in the Higgs
phase. In particular, at distances less than $\hxi=2\pi/\hk$ it looks like
there was a uniform external magnetic field. We can estimate that its
amplitude $B_{\rm avg}$ is
\begin{equation}
B_{\rm avg}^2\approx
\left\langle 
\int_0^{\hk}
\frac{d^3k}{(2\pi)^3}
B_i(\vec{k})
\int_0^{\hk}
\frac{d^3k'}{(2\pi)^3}
B_i(\vec{k}')
\right\rangle
\approx
\int_0^{\hk}\frac{d^3k}{(2\pi)^3}G_0(k)\sim T\hk^3.
\end{equation}
This magnetic flux must be confined into vortices, and 
because each vortex carries one flux quantum $\Phi_0=2\pi/e$, the
number density of vortices per unit area is
\begin{equation}
n\approx
\frac{e}{2\pi}B_{\rm avg}
\approx
\frac{e}{2\pi}T^{1/2}\hk^{3/2}.
\labx{equ:ndenspred}
\end{equation}

Even if we do not know $\tau(k)$, we can still make some concrete
predictions based on this scenario. For instance, the spatial
distribution of vortices turns out to be completely different than in
the KZ scenario. At distances shorter than $\hxi=2\pi/\hk$, 
the magnetic field
points in the same direction, which means that the vortices tend to be
aligned, while in the KZ scenario they prefer to be anti-aligned. This
prediction was confirmed in the simulations in Ref.~\cite{Hindmarsh:2000kd}.

\section{Classical simulations}
\labx{sect:classsim}
\subsection{Equilibrium}
\labx{sect:classeq}
In order to test the scenario presented in Sect.~\ref{sect:defect},
we carried out a number of numerical simulations. Let us first discuss
the simulations of the classical model (\ref{equ:eom}). 

Because we are interested in transitions that start from close to
thermal equilibrium, we will first have to thermalize the system.
This means preparing an ensemble of field configurations with
the canonical equilibrium distribution $\exp(-\beta H)$,
where $\beta=1/T$ and the Hamiltonian $H$ is
\begin{equation}
\labx{equ:clHam}
H=\int d^3x \left[
\frac{1}{2}\vec{E}^2+\frac{1}{2}(\vec\nabla\times\vec{A})^2
+\pi^*\pi+(D_i\phi)^*(D_i\phi)+m_T^2\phi^*\phi+\lambda(\phi^*\phi)^2
\right].
\end{equation}
Here $\pi=\partial_0\phi$ is the canonical momentum of $\phi$, and
in the temporal gauge ($A_0=0$), the electric field is simply 
$\vec{E}=-\partial_0\vec{A}$.
In addition, the fields must satisfy the Gauss law 
as an extra constraint
\begin{equation}
\vec{\nabla}\cdot\vec{E}=2e{\rm Im}\phi^*\pi.
\labx{equ:gauss}
\end{equation}

Because of the constraint (\ref{equ:gauss}), a straightforward Metropolis
algorithm would not work very well. Instead, we used a hybrid
algorithm, in which we thermalized the component of $\pi$ orthogonal to
Eq.~(\ref{equ:gauss}) with a heat bath algorithm, and performed a
number of Metropolis updates to the gauge field $\vec{A}$. Because
$\vec{A}$ does not appear in Eq.~(\ref{equ:gauss}), it leaves the
constraint unchanged. Then we evolved the system with the
equations of motion
\begin{eqnarray}
\labx{equ:fulleom}
\partial_0\vec{A}&=&-\vec{E},\nonumber\\
\partial_0\phi&=&\pi,\nonumber\\
\partial_0\vec{E}&=&\vec\nabla\times\vec\nabla\times\vec{A}
+2e{\rm Im}\phi^*D_i\phi,\nonumber\\
\partial_0\pi&=&D_iD_i\phi-m_T^2\phi-2\lambda(\phi^*\phi)\phi,
\end{eqnarray}
which, again, leave the Gauss law unchanged. We repeated this procedure
a number of times so that the system thermalized.

In our simulations, we used the couplings $e=0.3$ and $\lambda=0.18$. 
The lattice size was $120\times 120\times 20$ (the reason for choosing one
short dimension is discussed in Sec.~\ref{sect:classph}),
the lattice spacing was 
$\delta x=6T^{-1}$ and the time step was $\delta t=0.3T^{-1}$.
The details of the lattice implementation are given in 
Appendix~\ref{app:lattcl}.

\begin{figure}
\center
\begin{tabular}{ll}
a) & b)\cr
\epsfig{file=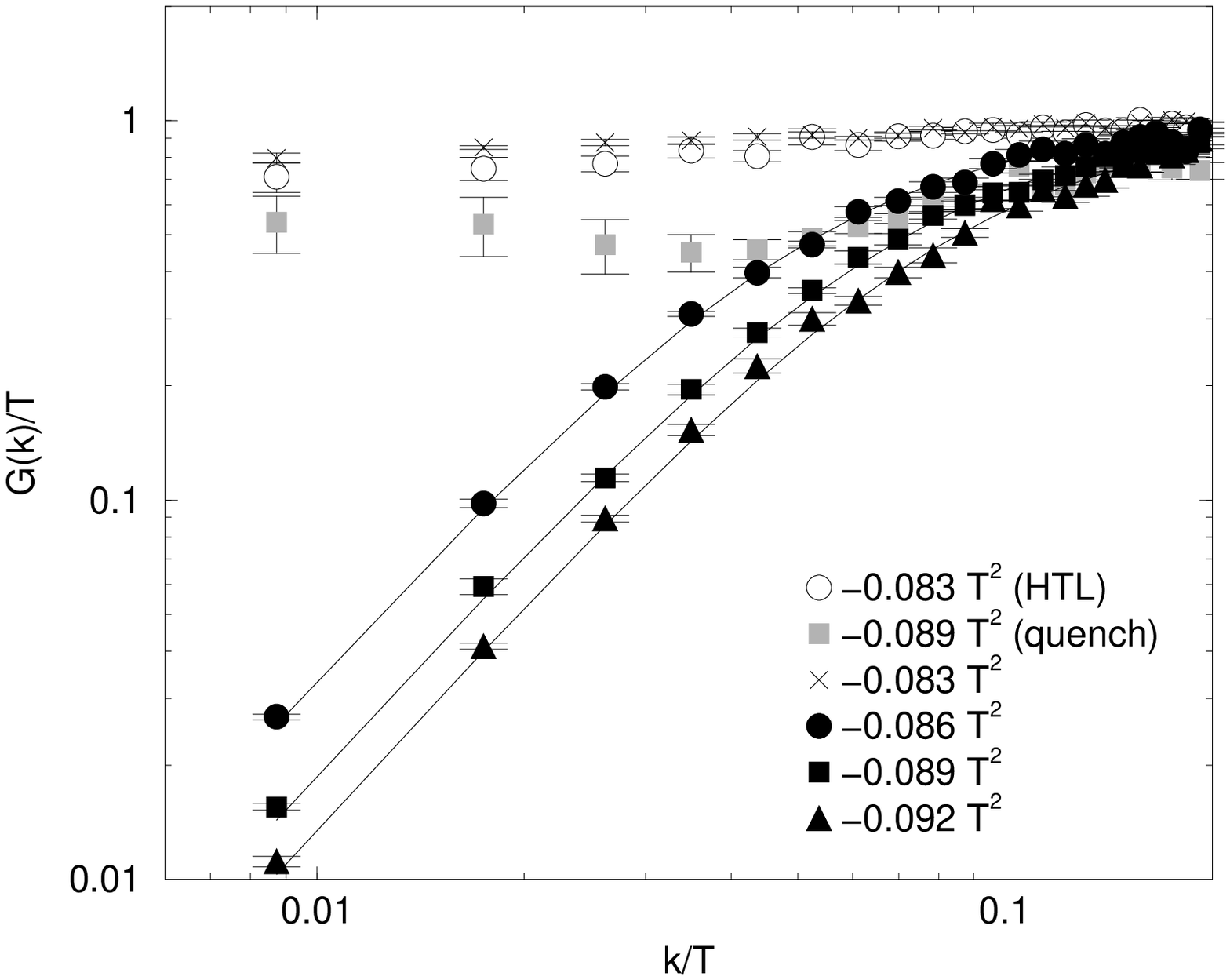,width=8cm,height=6cm}&
\epsfig{file=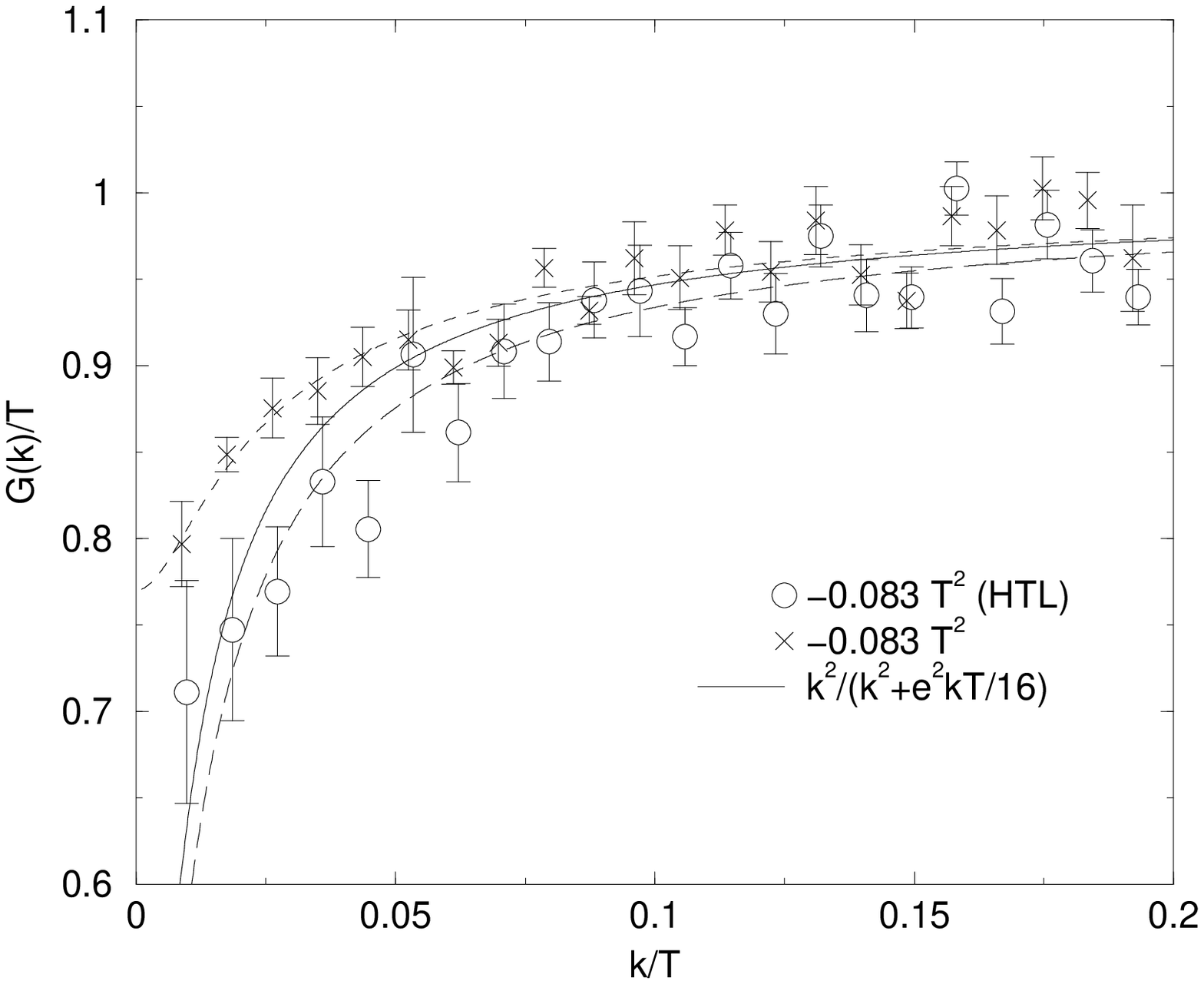,width=8cm,height=6cm}
\end{tabular}
\flushleft
\caption{
\label{fig:spatcorr}
The spatial correlator of the magnetic field in the Fourier space.
a) The solid lines are fits to Eq.~(\ref{equ:scpert})
and show that the system is
in the Higgs phase when $m^2\lsim-0.083T^2$.
The gray squares show the spatial correlator after a quench as
discussed in Section~\ref{sect:classph}. The open circles correspond to the
HTL simulations in Section~\ref{sect:htleq}.
b) The $m^2=-0.083T^2$ data on a linear scale. The short-dashed and
long-dashed lines show
the fits to Eqs.~(\ref{equ:full1loop}) and
(\ref{equ:nonpertselfenergy}), 
respectively, 
and the solid line shows the perturbative result (\ref{equ:pertknp}).
}
\end{figure}

Since $\lambda>e^2$, the phase transition is continuous. 
In order to determine the location of the transition point, and to
test the accuracy of the tree-level result (\ref{equ:scpert}), we
measured the correlator (\ref{equ:spatcorr}) at various values of
$m^2$, starting deep in broken phase. 
For thermalization to each value of $m^2$, we used 24 hybrid Monte
Carlo cycles each 
consisting of 400 Metropolis sweeps and 
time evolution for $\Delta t=600T^{-1}$. We carried out the
measurement in nine independently thermalized configurations,
measuring the average correlator in an unperturbed run of length
$\Delta t=12000T^{-1}$ (and $\Delta t=72000T^{-1}$ for $m^2=-0.083T^2$).
The results are shown in
Fig.~\ref{fig:spatcorr}a. The solid lines show that the 
agreement with Eq.~(\ref{equ:scpert}) is
excellent, and we find that the phase transition takes place at
$m^2\approx -0.083T^2$.

In the symmetric phase, the tree-level result (\ref{equ:scpert})
corresponds to a constant $G(k)=T$, but the data measured at
$m^2=-0.083T^2$ (see Fig.~\ref{fig:spatcorr}b), clearly
turns down at small momenta. 
It is customary to parameterize the corrections to the tree-level
result by introducing the static photon self energy $\Pi_T$,
defined by
\begin{equation}
G(k)=T\frac{k^2}{k^2+\Pi_T(k)}.
\end{equation}
Our measurements seem to contradict the results of
Kraemmer et al.~and Blaizot et
al.~\cite{Kraemmer:1995az,Blaizot:1995kg}, 
who showed that
after a resummation, the lowest-order term of $\Pi_T$ 
is proportional to $k^2$. Such a quadratic
term would only change the overall normalization of the correlator and
any higher-order terms would only modify the high-$k$ end of the
spectrum. However, they considered specifically the case in which the
zero-temperature Higgs mass vanishes, i.e., $m^2=0$. In that case, the
thermally generated effective mass for the Higgs field $M^2$,
which is approximately equal to $m_T^2$, is always $\sim
e^2T^2$ [see Eq.~(\ref{equ:mTren})] and 
acts as an infrared regulator in the loop
integral, making $\Pi_T(k)$ an analytic function of $k^2$. 
Because one can show that the constant term is forbidden, the
lowest-order term must be $O(k^2)$.

In our case, this argument breaks down because we are studying the
system at the transition point where the effective Higgs mass $M$
becomes
very small. 
In the momentum range 
$M \ll |\vec{k}| \ll eT$, the perturbative behaviour 
of $\Pi_T(k)$ at one loop order is 
\begin{equation}
\label{equ:pertknp}
\Pi_T(k)=\frac{e^2T}{16}k=0.005625Tk.
\end{equation}
This is shown as a solid line in Fig.~\ref{fig:spatcorr}b, and agrees
fairly well with the data. Assuming that perturbation theory is
applicable, the discrepancy at very low $k$ can be explained by a
non-zero effective Higgs mass, which arises because we are not exactly
at the transition point. With $M>0$, the one-loop result 
is \cite{Blaizot:1995kg}
\begin{equation}
\Pi_T(k)=\frac{e^2MT}{4\pi}\left[
\frac{4M^2+k^2}{2kM}\arctan\frac{k}{2M}-1\right].
\label{equ:full1loop}
\end{equation}
The best fit gives $M=(0.0040\pm0.0005)T$, and is shown 
in Fig.~\ref{fig:spatcorr}b as a short-dashed line.
It is curious how well it agrees with the measurements, because
one would expect perturbation theory to break down at low momenta
near the phase transition.

Assuming that the Higgs mass vanishes at the transition point,
all that remains is Eq.~(\ref{equ:pertknp}).
The existence of a linear term in the self energy $\Pi_T$ is
surprising, because it implies magnetic screening. This
screening is not as strong as it would be if there was a constant
term, in which case the correlator would fall exponentially in the
coordinate space. With a linear term, the low-momentum behaviour of
the transverse gauge field correlator is $\sim k^{-1}$, and
consequently, the long-distance behaviour in the coordinate space is
$\sim r^{-2}$ instead of the usual $\sim r^{-1}$.

We also measured real-time correlators in the same simulations.
In each of the nine independent configurations, we measured the
correlator $G(t,k)$ and took the average of the
results. Two examples, measured for $k=0.026T$ 
at $m^2=-0.083T^2$ and $m^2=-0.086T^2$, are
shown in Fig.~\ref{fig:realcorr}a. At all values of $m^2$ and $k$ we
measured, the data were well described by the function
\begin{equation}
G_{\rm fit}(t)=a_0\exp(-\gamma_Lt)+a_1\exp(-\gamma_p t)\cos(\omega_p
t+\delta),
\labx{equ:fiteq}
\end{equation}
where $a_0$, $a_1$, $\gamma_L$, $\gamma_p$, $\omega_p$ and $\delta$ are free
parameters. Physically, $\gamma_L$ is the Landau damping rate,
$\gamma_p$ is the plasmon damping rate, and $\omega_p$ is the plasmon
frequency.
We fitted this function to the data at each point and estimated the
errors using the bootstrap method. The results are shown in 
Figs.~\ref{fig:realcorr}b--d.

We can compare these results with perturbative calculations, but we
must keep in mind that since these simulations were carried out in a
classical lattice theory, the result is not the same as in quantum
theory in continuum. In the
symmetric phase, the plasmon frequency should behave as 
\begin{equation}
\omega_p(\k)=\sqrt{\k^2+m_p^2},
\end{equation}
where $m_p$ is the plasmon mass, which has been calculated in
classical lattice perturbation theory in 
Refs.~\cite{Arnold:1997yb,Bodeker:1998bt},
\begin{equation}
m_p^2\approx 0.086e^2\frac{T}{\delta x}\approx 0.0013T^2.
\end{equation}
We have plotted this curve in Fig.~\ref{fig:realcorr}b, and it agrees
very nicely with the measured frequencies at $m^2=-0.083T^2$. In the
Higgs phase, the photon becomes massive
due to the Higgs mechanism and this increases $m_p$.

The plasmon damping rate has not
been calculated perturbatively for the classical lattice theory. 
However, a calculation has been carried
out in the HTL-improved theory by Evans and
Pearson \cite{Evans:1997rw}, who showed that it was peaked just below the 
phase transition. Our classical results also show a rising trend as the
transition is approached, although we cannot directly compare the numerical
values.

\begin{figure}
\begin{tabular}{ll}
a)&b)\cr
\epsfig{file=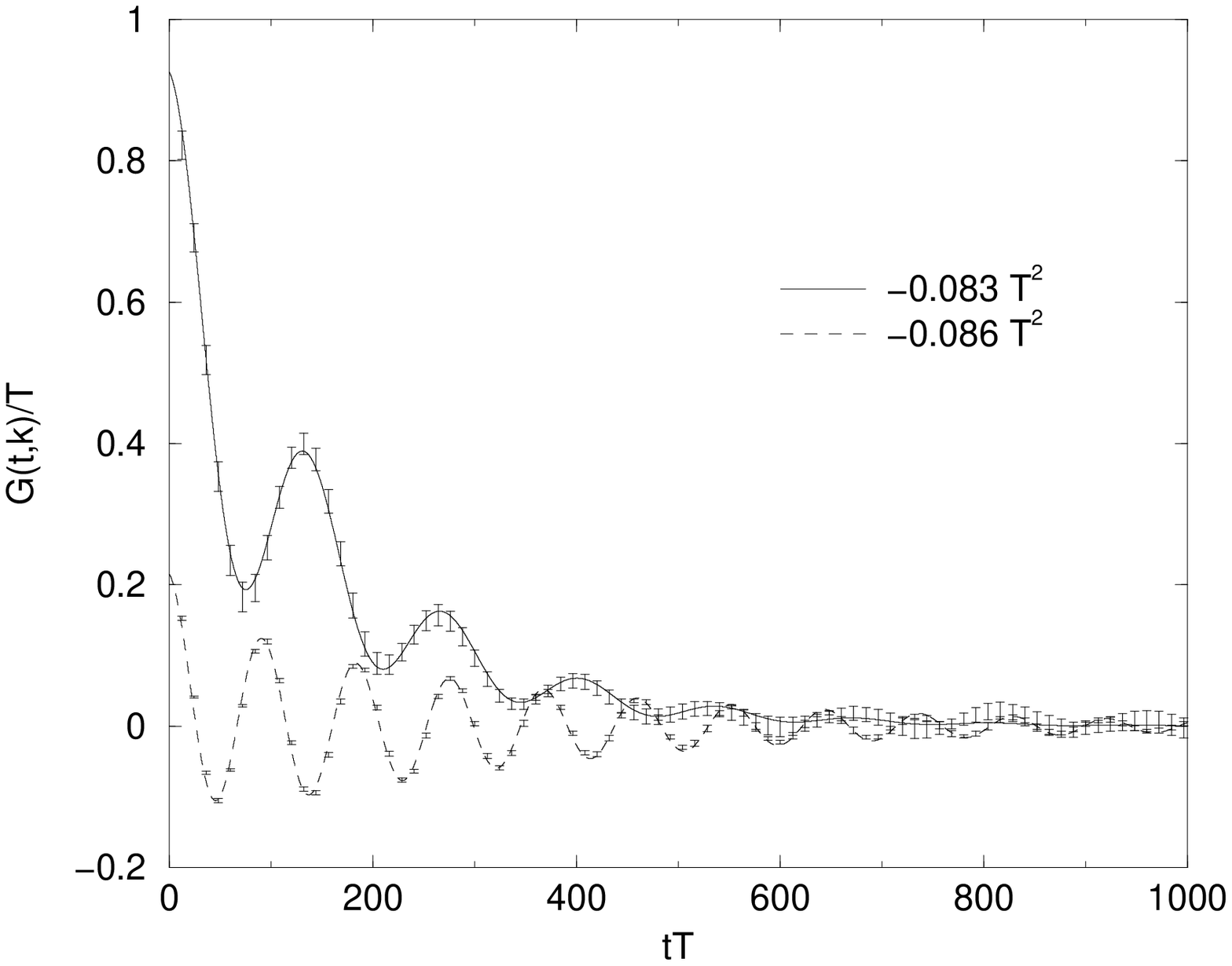,width=8cm,height=6cm}&
\epsfig{file=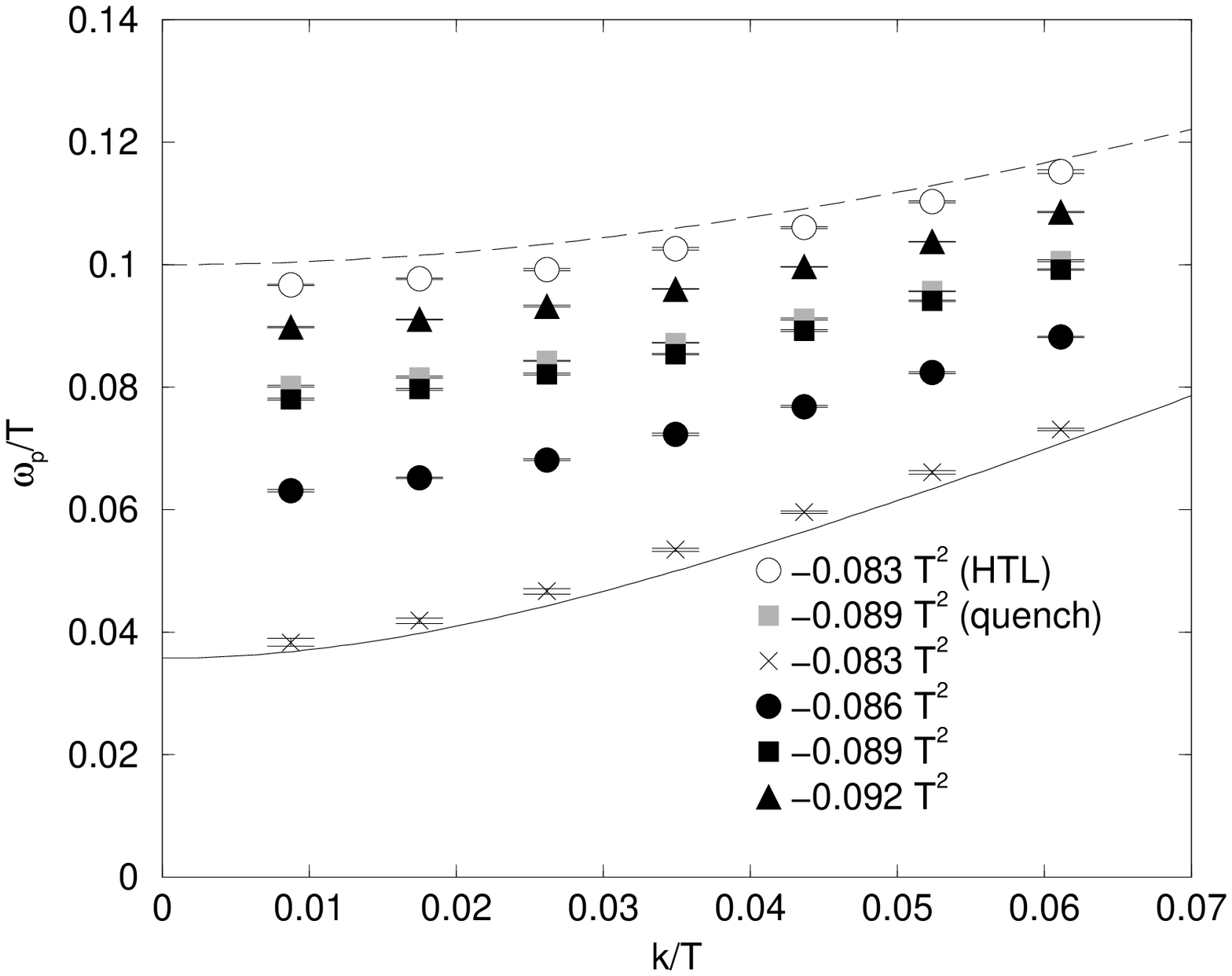,width=8cm,height=6cm}\cr
c)&d)\cr
\epsfig{file=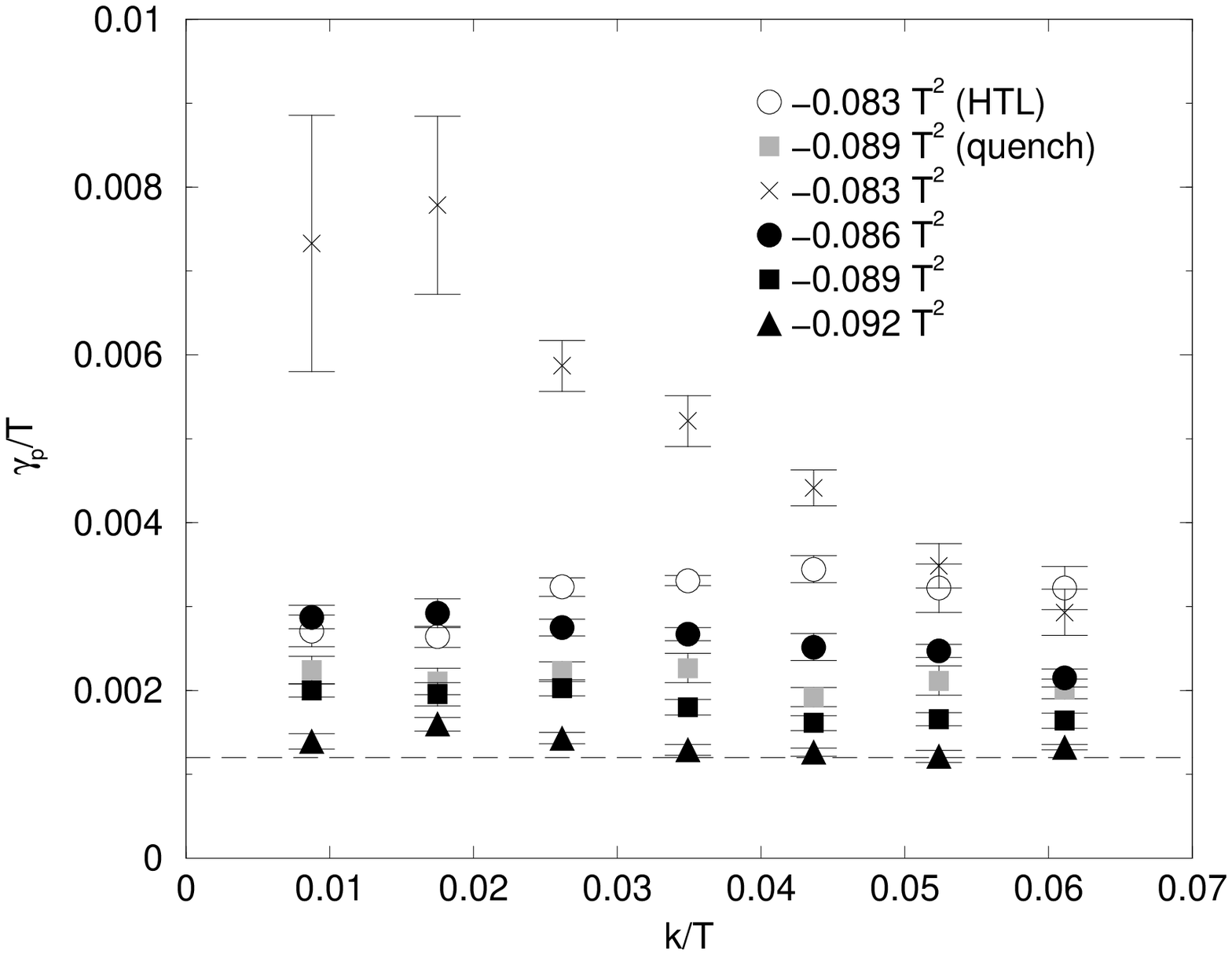,width=8cm,height=6cm}&
\epsfig{file=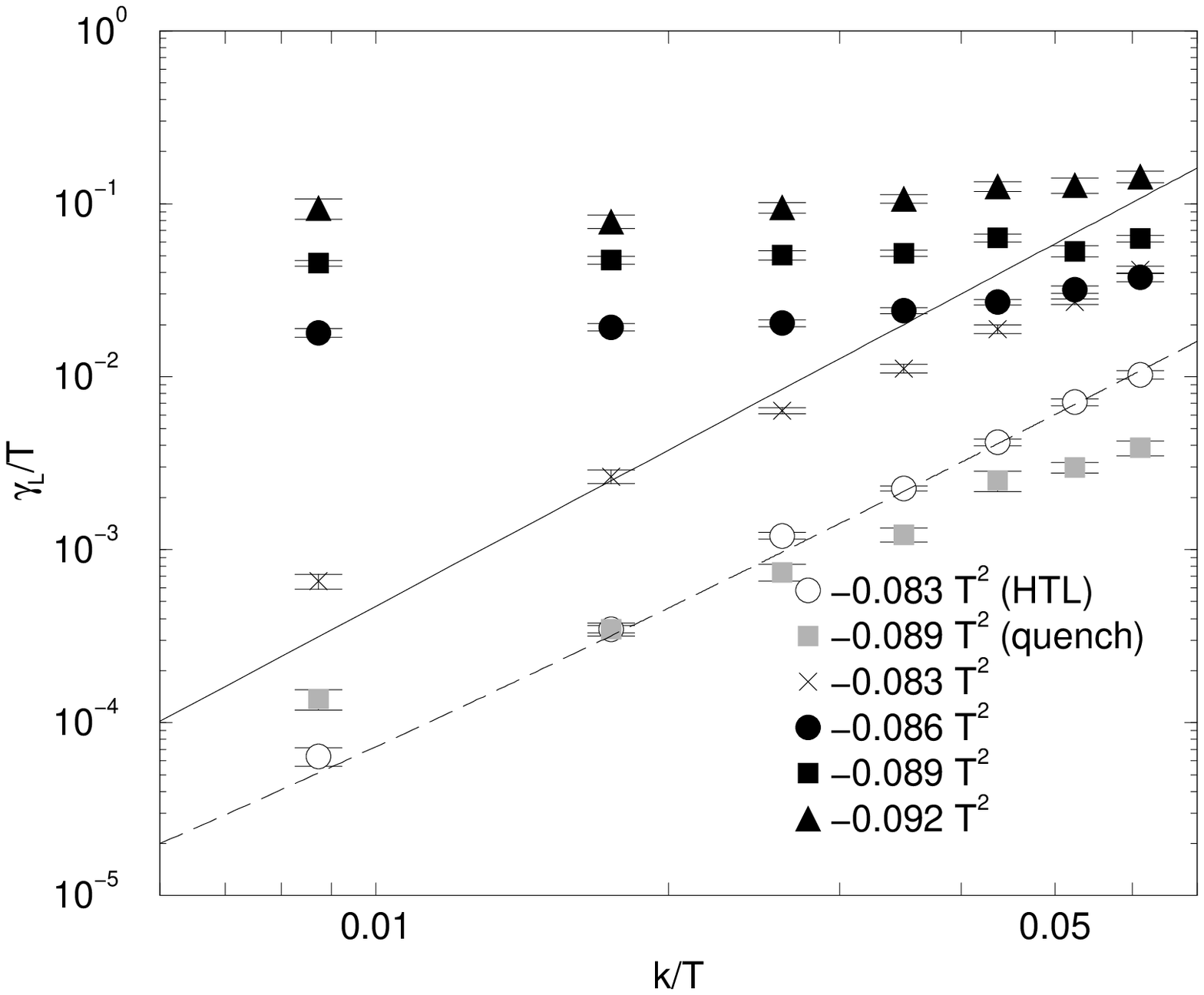,width=8cm,height=6cm}
\end{tabular}
\caption{
\labx{fig:realcorr}
The real-time correlators measured at different values of $m^2$.
a) Two examples of real-time correlators, measured at $m^2=-0.083T^2$
(solid)
and $m^2=-0.086T^2$ (dashed),
$k=0.026T$, together with fits of the form (\ref{equ:fiteq}).
b) The fitted plasmon frequencies $\omega_p$, and the perturbative estimates.
c) The fitted plasmon damping rates $\gamma_p$.
d) The fitted Landau damping rates $\gamma_L$, and the perturbative estimates.
In plots b--d, the gray squares correspond to the state of the system
after a quench and open circles to the HTL simulations discussed in
Section~\ref{sect:htleq}.
}
\end{figure}

The Landau damping rate in the symmetric phase has been
calculated perturbatively in Ref.~\cite{Arnold:1997yb},
\begin{equation}
\gamma_L\approx \frac{7\delta x}{e^2T}k^3\approx 470\frac{k^3}{T^2}.
\end{equation}
However, in our results the dependence on $\k$ is milder and we find
the best fit with
$\gamma_L\approx 15T^{-1.1}k^{2.1}$ above the transition point.
Below the transition point, the exponential decay rate $\gamma_L$
becomes very large so that at late times, the correlator simply
oscillates around zero (see Fig.~\ref{fig:realcorr}a).
In Fig.~\ref{fig:realcorr}d, 
this behaviour can be seen in the values of the Landau
damping rate.

In the defect formation scenario, the freeze-out occurs when
Eq.~(\ref{equ:adiabatic}) ceases to be satisfied. This happens most
easily at the transition point, and there the real-time correlators
are still the same as in the symmetric phase. Therefore, it is the
symmetric phase correlators that determine the relevant time scale
$\tau(k)$, and Fig.~\ref{fig:realcorr} shows clearly that for the
long-wavelength modes, the lowest time scale is that of Landau
damping, $\tau(k)\approx\gamma_L(k)^{-1}$. 

Substituting $\tau(k)=\gamma_L(k)^{-1}$ to
Eq.~(\ref{equ:adiabatic}), we find
\begin{equation}
\frac{1}{\hk^2}\frac{\delta m^2}{\tau_Q}\approx \gamma_L\approx
15T^{-1.1}\hk^{2.1},
\end{equation}
which implies
\begin{equation}
\hk\approx\left(\frac{0.0061}{T\tau_Q}\right)^{1/4.1}T
\approx 0.29T (T\tau_Q)^{-0.24}\propto\tau_Q^{-0.24}.
\labx{equ:classpred}
\end{equation}

\subsection{Phase transition}
\labx{sect:classph}
We simulated the phase transition by thermalizing a number of
configurations with $m^2=m_0^2=-0.044T^2$, and solving numerically
the equations of motion (\ref{equ:fulleom}), 
varying the mass term with time according to
\begin{equation}
m^2(t)=m_0^2-\delta m^2\left(\frac{4}{3\pi}\arctan(t/\tau_Q-1)+\frac{1}{3}
\right),
\labx{equ:masschange}
\end{equation}
where $\delta m^2=0.089T^2$.
This form of $m^2(t)$ has the advantage that long time after the transition,
the system is in thermal equilibrium, making it possible
to compare the final states.

The 
vortices produced in the transition are closed loops, and in general
they will soon shrink to a point and disappear, which makes it very difficult
to even define what we mean by the final vortex number. 
However, on a periodic lattice some of the vortex loops can be 
non-contractible, i.e.~wind around the lattice in some direction. These
loops can only disappear if they annihilate with another vortex of
opposite direction, but this process is very slow, since the
interactions between the vortices are exponentially suppressed at
long distances. Therefore we chose one of the lattice dimensions much
shorter than the other two (but still much longer than 
the microscopic length scales such as the Debye screening length
$m_D^{-1}$.) A long time after the transition, when
the system is deep in the broken phase, these vortices still remain and
are well-defined macroscopic objects, and they can be counted
without any ambiguities using the gauge-invariant
lattice definition for the winding
number~\cite{Ranft:1983hf,Kajantie:1998bg}.

Because the scenario of defect formation discussed here is based on
the assumption that the distribution of the magnetic field freezes in
the transition, we can carry out a very simple test for the scenario
by quenching the system through the transition and measuring the
spatial correlator. We used $\tau_Q=300T^{-1}$ and stopped the quench
at $m^2=-0.089T^2$ to carry out the measurement. 
The spatial correlator is shown in
Fig.~\ref{fig:spatcorr}a as gray squares, 
and indeed, resembles much more the symmetric
phase correlator than the equilibrium correlator at the same value of
$m^2$. We also measured the real-time correlator, and the
corresponding time scales are shown in
Figs.~\ref{fig:realcorr}b--c. The plasmon frequency and decay rate do
not differ significantly from their equilibrium values, but Landau
damping gets extremely slow. This means that once a mode has fallen out
of equilibrium at the transition point, it takes a very long time
before it thermalizes.

When one of the dimensions is shorter than $\hxi=2\pi/\hk$, the prediction
(\ref{equ:ndenspred}) changes into~\cite{Hindmarsh:2000kd}
\begin{equation}
n\approx
\frac{e}{2\pi}T^{1/2}L_z^{-1/2}\hk^{-1}.
\labx{equ:classdens}
\end{equation}
We tested this prediction by simulating the time evolution with
different values of $\tau_Q$ and measuring the vortex number a long time
after the transition
at $t=\tau_Q+2400T^{-1}$. The results were published in
Ref.~\cite{Hindmarsh:2000kd}, 
and are shown in Fig.~\ref{fig:taudep}. Each point is an average over
around 15 runs starting from different initial configurations. 

\begin{figure}
\begin{center}
\epsfig{file=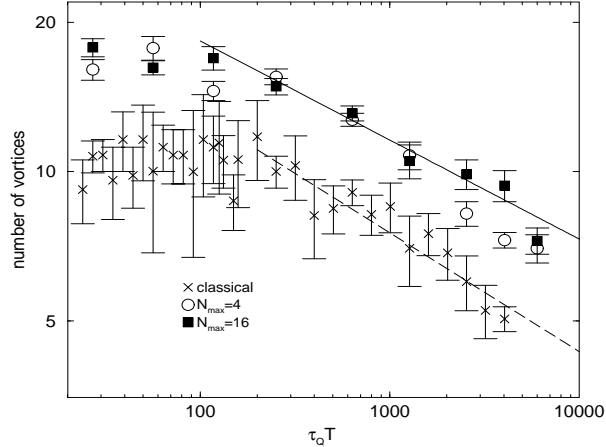,width=8cm,height=6cm}
\end{center}
\caption{
The dependence of the final number of vortices on the quench rate
$\tau_Q$. 
The dashed line is a power-law fit of the data at $\tau_Q>200T^{-1}$ 
with the exponent $-0.24$ predicted by
Eq.~(\ref{equ:classpred}).
The circles and squares correspond to the HTL simulations discussed in
Section~\ref{sect:htlph}, and the solid line is a power-law fit
of $\tau_Q>100T^{-1}$
using the exponent $-0.2$ predicted by Eq.~(\ref{equ:htlpred}).
}
\labx{fig:taudep}
\end{figure}

Combining the result $\hat{k}\sim\tau_Q^{-0.24}$ from
Eq.~(\ref{equ:classpred}) with Eq.~(\ref{equ:classdens}), we expect
$N\approx c\tau_Q^{-0.24}$. If we determine the constant $c$ from the
best fit to the data at $\tau_Q>200 T^{-1}$, 
we find $c=39.5\pm 1.0$ and
$\chi^2=7.6/13$ dof. This is shown in Fig.~\ref{fig:taudep} as the
dashed line.
If we also leave the exponent as a free parameter,
we find $N=(43.9\pm 7.9)(T\tau_Q)^{-0.255\pm0.026}$, with
$\chi^2=7.2/12$ dof. 

If we calculate precisely the prediction (\ref{equ:classdens}) using
Eq.~(\ref{equ:classpred}), we find $c\approx 650$ for the above fit
parameter. However, it is not surprising that it differs from the
measured value by a factor of $\approx 15$, 
not only because have we neglected factors or $2\pi$ and other numerical
factors, but also because in our estimates we have assumed that the
mechanism is ideally efficient and that all the magnetic flux is
converted into vortices. 

From the cosmological point of view, we are more interested in the
area density of vortices 
in a fully three-dimensional case than in a
thin box. Because the topology of the system does not prevent vortex
loops from shrinking, the resulting network is not stable, unless it
is stabilized by the expansion of the universe. We can therefore only
use Eq.~(\ref{equ:ndenspred}) as an estimate for the area density
of
vortices immediately after the transition.
Comparing 
Eqs.~(\ref{equ:ndenspred}) and (\ref{equ:classdens}), we find
\begin{equation}
n_{\rm 3D}=\left(\frac{2\pi}{e}\right)^{1/2}T^{-1/4}L_z^{3/4}n_{\rm 2D}^{3/2},
\labx{equ:convert3D}
\end{equation}
where $n_{\rm 2D}$ is given by Eq.~(\ref{equ:classdens})
and $n_{\rm 3D}$ is the three-dimensional area density given by 
Eq.~(\ref{equ:ndenspred}).
In our case,
$L_z=120T^{-1}$ and $A=5.2\times10^{5}T^{-2}$, and we find
\begin{equation}
n_{\rm 3D}\approx 166T^{-1}n_{\rm 2D}^{3/2}\approx
1.3\times 10^{-4}(T\tau_Q)^{-0.38}.
\end{equation}

\section{HTL simulations}
\labx{sect:htlsim}

\subsection{Equilibrium}
\labx{sect:htleq}
As discussed in Section~\ref{sect:ahm}, the classical theory discussed
above does not describe the dynamics of the quantum theory
correctly. Therefore we also carried out the same simulations with the
HTL improved theory (\ref{equ:lagrHTL}).
In the same way as in Section~\ref{sect:classeq}, we first studied the
equilibrium properties of the theory. 
We used the same couplings $e=0.3$ and $\lambda=0.18$
as in the classical case, and the same lattice. 
With these parameters, the Debye mass has the value
$m^2_D=0.03T^2$. The number of Legendre modes was $N_{\rm max}=4$.
The only effect of the HTL
corrections to the thermodynamics is to give an extra contribution to
electric screening, so we expect that the phase diagrams of the
theories are practically the same. Therefore we only carried out the
equilibrium measurements at the transition point, $m^2=-0.083T^2$.

Using the Hamiltonian (\ref{equ:ham}), we first prepared the thermal
initial conditions
with a Monte Carlo algorithm. This can be done in two steps:
\begin{enumerate}
\item{Generate the soft mode configuration.
The part of the Hamiltonian that involves the hard modes is Gaussian
and therefore they can be integrated out exactly. This results
in a theory with only the soft fields $\phi$ and $A_i$ and their
canonical momenta. The Gauss law leads to
an extra Debye screening term 
\begin{equation}
\delta H_{\rm Debye}=\frac{1}{2m_D^2}
\left(
\vec{\nabla}\cdot\vec{E}-2e{\rm Im}\phi^*\pi
\right).
\labx{equ:debscreen}
\end{equation}
in the Hamiltonian.\footnote{In principle, the canonical momenta could also
be integrated out at this stage, resulting in a theory with
an extra neutral scalar field $A_0$. This would lead to an effective
theory that is equivalent with dimensional reduction at one-loop
order~\cite{Ginsparg:1980ef}.}
We used a Metropolis algorithm for $\phi$ and a heat bath algorithm for
all other fields, and
carried out around 10000 thermalization sweeps.}
\item{
Generate the hard modes in the background of the soft
modes generated in step 1. 
Since the Hamiltonian is Gaussian it could in principle be
diagonalized and the field values could be taken directly from a 
normal distribution. However, we did this only for $\theta$ and its
momentum $\Pi$. Note that the lowest Legendre mode of $\Pi$ is
fixed by the Gauss law,
\begin{equation}
\Pi^{(0)}=-\frac{1}{m_D}\left(\vec{\nabla}\cdot\vec{E}-2e{\rm Im}
\phi^*\pi
\right)
.
\end{equation}
The field $\vec{f}$ and its momentum $\vec{F}$ were generated using a
heat bath algorithm with around 5000 sweeps.}
\end{enumerate}

Again, we evolved the configurations taken from the thermal ensemble
using the equations of motion (\ref{equ:eomLeg}) for the time 
$\Delta t=12000T^{-1}$, measuring the equal-time and real-time correlators.
The equal-time correlator is shown as open circles in 
Fig.~\ref{fig:spatcorr}, 
and it
agrees fairly well with the corresponding classical
correlator. This time, the fit to the perturbative one-loop result
(\ref{equ:full1loop}) 
favours $M=0$, which suggests that because of the presence of the hard
modes,
the transition point is at slightly larger $m^2$ than in the
classical case, and that
$m^2=-0.083T^2$ is very close to it.

Because perturbation theory cannot be trusted at low
momenta, we do not adopt the perturbative result (\ref{equ:pertknp}),
but instead we
simply assume that the self energy is linear at small momenta
\begin{equation}
\label{equ:nonpertselfenergy}
\Pi_T(k)=k_{\rm np}k,
\end{equation}
and determine the coefficient $k_{\rm np}$
from the best fit to the data,
which gives
\begin{equation}
k_{\rm np}=(0.0071\pm0.0005)T,
\end{equation}
and use this in our later estimates.

The hard modes have a significant effect on the real-time
correlators, shown in Figs.~\ref{fig:realcorr}b--d. Because they mimic
the effect of the hard modes in the continuum quantum theory, we should now be
able to compare the results with the standard perturbative
calculations. For the plasmon mass, the perturbative result 
is~\cite{ref:PhysKin,Kraemmer:1995az}
\begin{equation}
m_p=\frac{1}{\sqrt{3}}m_D=0.1T.
\end{equation}
The dashed line in Fig.~\ref{fig:realcorr}b shows the corresponding
curve
$\omega_p=\sqrt{k^2+m_p^2}$. The measured values are slightly below
this curve, but the agreement is still very good.

The continuum plasmon damping rate at zero momentum has
been computed perturbatively in the HTL approximation
in Refs.~\cite{Kraemmer:1995az,Evans:1997rw}, 
\begin{equation}
\gamma_p(T) =
\frac{e^2T}{24\pi}A(T/T_c,\lambda/e^2).
\end{equation}
The function $A(T/T_c,\lambda/e^2)$ 
takes the value 1 at the critical point,
peaks at value $\sim 1$ just
below it, and vanishes above it when 
$m_p > 2M$. In our case, the damping rate at the transition
point would be
\begin{equation}
\gamma_p(T)\approx 0.0012T,
\end{equation}
which is shown in Fig.~\ref{fig:realcorr}c as a dashed line and 
is lower than value the measured at $m^2=-0.083T^2$ by roughly a factor of
three. It has been observed earlier \cite{Bodeker:1998bt}, that also
in the SU(2) theory, the perturbative calculation underestimates the
plasmon damping rate significantly.
We can also note that the damping rate is lower than in the classical
theory.

The Landau damping rate can be obtained directly from the HTL improved
Lagrangian and is~\cite{ref:PhysKin}
\begin{equation}
\gamma_L=\frac{4k^2}{\pi
m_D^2}(k+k_{\rm np})\approx42.4\frac{k^2}{T^2}(k+k_{\rm np}),
\end{equation}
where we have taken into account the linear contribution
(\ref{equ:nonpertselfenergy})
to the self energy. 
This curve is shown as the dashed line in Fig.~\ref{fig:realcorr}d, and
agrees well with the measured values.
Again, we can conclude that the relevant time scale is that of Landau
damping. If we can ignore the linear correction, we find 
[cf.~Eq.~(\ref{equ:classpred})]
\begin{equation}
\hk\approx\left(\frac{\pi}{4}m_D^2\frac{\delta
m^2}{\tau_Q}\right)^{0.2}\approx0.29T(T\tau_Q)^{-0.2}
\sim \tau_Q^{-0.2},
\labx{equ:htlpred}
\end{equation}
and in slowest transitions, we have
\begin{equation}
\hk\approx\left(\frac{\pi}{4}\frac{m_D^2}{k_{\rm np}}\frac{\delta
m^2}{\tau_Q}\right)^{0.25}\approx0.74T(T\tau_Q)^{-0.25}
\sim \tau_Q^{-0.25}.
\labx{equ:htlpred2}
\end{equation}
However, this corresponds to $\tau_Q\gg 10^8T^{-1}$, which is much
larger than the values of $\tau_Q$ we are able to use in out simulations.

\subsection{Phase transition}
\labx{sect:htlph}
As in Sect.~\ref{sect:classph}, we studied the non-equilibrium
dynamics of the phase transition by starting from thermal
configurations at $m^2=-0.044T^2$ and evolving the system with a
time-dependent mass term (\ref{equ:masschange}). We used two different
values for $N_{\rm max}$, $4$ and $16$.

In Fig.~\ref{fig:taudep}, we show how the final vortex number, measured
at $t=2\tau_Q+2400T^{-1}$ depends on the quench rate $\tau_Q$.
Each data point is an average over 20--30 runs. 
In fast transitions, $\tau_Q\lsim 1000T^{-1}$, the results for $N_{\rm
max}=4$ and $16$ agree, but in slower transitions there is a
statistically significant difference. This can be understood in terms
of Eq.~(\ref{equ:crittime}): For small $\tau_Q$, $N_{\rm max}=4$ is
sufficient because by the time the approximation breaks down, the
system is already so deep in the Higgs phase that the vortex number
cannot change any more. When $\tau_Q$ gets larger, eventually a point
is reached at which the breakdown occurs so early that it would have
an effect on the final state. This may have happened even for $N_{\rm
max}=16$ in the slowest transitions with $\tau_Q=6000T^{-1}$.

Combining Eq.~(\ref{equ:htlpred}) with Eq.~(\ref{equ:classdens}), 
we find $N\sim\tau_Q^{-0.2}$. A fit to the $N_{\rm max}=16$ 
data at $\tau_Q>100T^{-1}$
with $N=c(T\tau_Q)^{-0.2}$ gives $c=45.5\pm0.9$ with $\chi^2=6.8/6$ dof,
which shows that the results are compatible with the prediction.
The fit is shown in Fig.~\ref{fig:taudep} as a solid line.
The prediction of Eq.~(\ref{equ:classdens}) is $c\approx
660$, which is greater than the measured value by a factor of
$\approx 15$. This is the same factor found
in the classical case, which further supports our scenario.
If we keep the exponent as a free parameter, we find
$N=(45.9\pm 4.6)(T\tau_Q)^{-0.201\pm0.015}$, with $\chi^2=6.8/5$ dof.

Again, we relate the results to three dimensions using
Eq.~(\ref{equ:convert3D}), and find
\begin{equation}
n_{\rm 3D}\approx 1.4\times10^{-4}(\tau_QT)^{-0.30}.
\end{equation}
The conjectured non-perturbative behaviour at low momenta
(\ref{equ:nonpertselfenergy}) would change this scaling law in very
slow transitions. Eq.~(\ref{equ:ndenspred}) would become
$n\propto\hk^2$, and Eq.~(\ref{equ:htlpred2}) would therefore imply
$n\propto\tau_Q^{-0.5}$.

\section{Conclusions}
\label{sect:conc}

In this paper we have presented a thorough study of the dynamics of the
high temperature phase transition in the Abelian Higgs model,   
using both classical and HTL improved approximations.  
The aims were twofold. The first was to measure the equilibrium properties of
the theory, and in particular find the relevant timescale for the decay of the
long wavelength modes of the gauge field, which were identified in
\cite{Hindmarsh:2000kd} as crucial to the understanding of vortex formation. 

In Ref.~\cite{Hindmarsh:2000kd}, 
we performed numerical simulations of the Abelian
Higgs model phase transition in real time by using the classical
theory and changing the mass parameter of the
Higgs field over a characteristic quench time $\tau_Q$.  In this
paper,
our second aim was to
do the same quenches with the HTL improved theory, and to compare the
resulting scaling law for the number of vortices $N$ formed as a
function of the quench time.

Our measurements of the equilibrium correlators show that perturbation
theory gives a reasonable 
account of their behaviour, except perhaps for the plasmon damping rate.
A new and unexpected result is that 
near the transition, the equal-time correlator exhibits power-law
magnetic screening, with a coefficient that is similar but not equal
to the perturbative one.
In the HTL improved
simulations, the numerical values
of the plasmon mass and the Landau damping rate agree well
with the perturbative values in the Coulomb phase. 
For the plasmon damping rate, however, the measured value was significantly
higher than the perturbative estimate.
As might be expected, the
agreement is not as good in the classical theory: 
although the plasmon mass agrees well, the
dependence of the Landau damping rate $\gamma_L$ on wavenumber $k$ is
$\gamma_L \sim k^{2.1}$
rather than the expected $k^3$. 

We found significant differences between the scaling laws for the classical
simulations, and for HTL improved quenches with $N_{\rm max}=4$ and 
$N_{\rm max}=16$.  The difference between the classical scaling law and the HTL
improved one with $N_{\rm max}=16$ can be ascribed to the discrepancy in the
Landau damping rate, and lends force to the contention made in 
\cite{Hindmarsh:2000kd} that it is the balancing of the cooling rate with the
Landau damping rate which decides the length scale above which the fields fall
out of equilibrium.  The difference between $N_{\rm max}=4$ and 
$N_{\rm max}=16$ can be understood as stemming from a lack of phase space in
the hard modes, which causes the HTL approximation to break down at times
greater than $t_c(k) = 4N_{\rm max}/k$.

Our simulations have been carried out to leading order in the couplings $e$ and
$\lambda$, and hence do not include the effect of high momentum
transfer scattering between the hard modes. However, the hard
modes can still scatter by exchanging soft modes, 
we do not expect this
to change the dynamics qualitatively on the  
relatively short time scales which we have been able to study.
In very slow transitions, it may be that hard mode 
scattering and also non-perturbative effects in the photon self-energy start
to become important and
and the predicted scaling law $n\propto\tau_Q^{-0.3}$ ceases to be
valid.  

\acknowledgments
We would like to thank Tim Evans, Edmond Iancu and Mikko Laine
for useful discussions.
Part of this work was conducted on the SGI Origin platform using COSMOS
Consortium facilities, funded by HEFCE, PPARC and SGI.
We also acknowledge computing support from
the Sussex High Performance Computing Initiative.
AR was supported by PPARC and also partly by the University of Helsinki.
\appendix

\section{HTL improved equations of motion}
\label{app:htleom}
In Sect.~\ref{sect:ahm} the local formulation of the HTL improved Abelian Higgs
model was described.  In this Appendix we detail the resulting equations of
motion for the fields $\phi$ and $\vec{A}$ and
Legendre modes $\theta$ and
$\vec{f}$, which encode the effect of high
momentum ($k \gsim T$) particles. 

These fields satisfy the equations of motion
\labx{equ:formulation}
\begin{eqnarray}
\partial_0^2{\vec{A}}&=&-\vec\nabla\times\vec\nabla\times\vec{A}
-2e{\rm Im}\phi^*\vec{D}\phi\nonumber\\
&&
+m_D\int_0^1dzz^2\left(
\vec\nabla\theta-m_D\vec{A}+
\sqrt{\frac{1-z^2}{2z^2}}\vec\nabla\times\vec{f}
\right),
\nonumber\\
\partial_0^2{\vec{f}}(z)&=&-z^2\vec{\nabla}\times\vec{\nabla}\times
\vec{f}+m_Dz\sqrt\frac{1-z^2}{2}
\vec\nabla\times\vec{A},
\nonumber\\
\partial_0^2{\theta}(z)&=&z^2\vec{\nabla}\cdot\left(
\vec{\nabla}\theta-m_D\vec{A}\right).
\end{eqnarray}
Since these equations for $\vec{f}$ and $\theta$ 
are linear, it is easy to solve them and
show that the dynamics of $\phi$ and $A_i$ is identical to the
original non-local theory (\ref{equ:nonloc}).

Not only is this reformulation of the theory local, but the
equations of motion are in a canonical form and we can therefore
write down the corresponding Hamiltonian
\begin{eqnarray}
H&=&\int d^3x\int_0^1dz\Biggl[
\frac{1}{2}\vec{E}^2+\frac{1}{2}(\vec\nabla\times\vec{A})^2
+\pi^*\pi+(D_i\phi)^*(D_i\phi)+m_T^2\phi^*\phi+\lambda(\phi^*\phi)^2
\nonumber\\&&
+\frac{1}{2}\vec{F}^2+\frac{1}{2}{\Pi}^2+
\frac{z^2}{2}(\vec\nabla\times\vec{f})^2
+\frac{z^2}{2}(\vec{\nabla}\theta-m_D\vec{A})^2
-m_Dz\sqrt\frac{1-z^2}{2}\vec{f}\cdot\vec\nabla\times\vec{A}
\Biggr],
\labx{equ:hamf}
\end{eqnarray}
where $\vec{F}=\partial_0{\vec{f}}$ and $\Pi=\partial_0\theta$ 
are the canonical momenta of $\vec{f}$ and $\theta$,
respectively. We also need two extra conditions, namely the transverseness of
$\vec{f}$ and Gauss's law
\begin{eqnarray}
\vec\nabla\cdot\vec{f}&=&\vec\nabla\cdot\vec{F}=0,
\labx{equ:transverse}\\
\vec\nabla\cdot\vec{E}&=&-m_D\int_0^1dz\Pi(z)+2e{\rm Im}\phi^*\pi.
\labx{equ:gaussHTL}
\end{eqnarray}

The $z$ dependence of $\vec{f}$ and $\theta$ is discretized 
\cite{Rajantie:1999mp} by introducing the Legendre modes
\begin{eqnarray}
\vec{f}^{(n)}&=&\int_0^1dzz\sqrt\frac{2}{1-z^2}P_{2n}(z)\vec{f}(z),
\quad
\theta^{(n)}=\int_0^1dzP_{2n}(z)\theta(z),
\nonumber\\
\vec{F}^{(n)}&=&\int_0^1\frac{dz}{z}\sqrt\frac{2}{1-z^2}P_{2n}(z)\vec{F}(z),
\quad
\Pi^{(n)}=\int_0^1dzP_{2n}(z)\Pi(z).
\labx{equ:defLeg}
\end{eqnarray}
Note that we have used slightly different definitions 
from Ref.~\cite{Rajantie:1999mp}, in order to write the Hamiltonian in
a practical form.

In terms of these modes, the equations of motion become
\begin{eqnarray}
\partial_0\vec{A}&=&-\vec{E},\nonumber\\
\partial_0{\vec{f}}^{(n)}&=&C^+_n\vec{F}^{(n+1)}
+C^0_n\vec{F}^{(n)}
+C^-_n\vec{F}^{(n-1)},
\nonumber\\
\partial_0{\theta}^{(n)}&=&\Pi^{(n)},
\nonumber\\
\partial_0{\vec{E}}&=&\vec\nabla\times\vec\nabla\times\vec{A}
+2e{\rm Im}\phi^*\vec{D}\phi+\frac{1}{3}m_D^2\vec{A}
\nonumber\\&&
-\frac{1}{3}m_D\left(
\vec\nabla\theta^{(0)}+2\vec\nabla\theta^{(1)}+
\vec\nabla\times\vec{f}^{(0)}-\vec\nabla\times\vec{f}^{(1)}
\right),
\nonumber\\
\partial_0\vec{F}^{(n)}&=&-\vec{\nabla}\times\vec{\nabla}\times
\vec{f}^{(n)}+m_D\vec{\nabla}\times\vec{A}\delta_{n,0}
,\nonumber\\
\partial_0\Pi^{(n)}&=&
C^+_n\vec\nabla^2\theta^{(n+1)}
+C^0_n
\vec\nabla^2\theta^{(n)}
+C^-_n\vec\nabla^2\theta^{(n-1)}-m_D\vec\nabla\cdot\vec{A}
\left(\frac{1}{3}\delta_{n,0}+\frac{2}{15}\delta_{n,1}
\right),
\labx{equ:eomLeg}
\end{eqnarray}
where
\begin{equation}
C^+_n=\frac{(2n+1)(2n+2)}{(4n+1)(4n+3)},\quad
C^0_n=\frac{1}{4n+1}\left(\frac{(2n+1)^2}{4n+3}+\frac{4n^2}{4n-1}\right),\quad
C^-_n=\frac{2n(2n-1)}{(4n+1)(4n-1)}.
\end{equation}
The Hamiltonian becomes
\begin{eqnarray}
H&=&\int d^3x\Biggl\{
\frac{1}{2}\vec{E}^2+\frac{1}{2}(\vec\nabla\times\vec{A})^2+
\pi^*\pi+(D_i\phi)^*(D_i\phi)+m_T^2\phi^*\phi+\lambda(\phi^*\phi)^2
\nonumber\\&&
+\sum_{n=0}^\infty
\left[
\frac{1}{4}\frac{1}{4n+3}
\left((2n+1)\vec{F}^{(n)}+(2n+2)\vec{F}^{(n+1)}\right)^2\right.
\nonumber\\&&
-\frac{4n+1}{4}
\left(C^+_n\vec{F}^{(n+1)}+C^0_n\vec{F}^{(n)}
+C^-_n\vec{F}^{(n-1)}\right)^2
\nonumber\\&&
+\frac{4n+1}{4}\left(\vec{\nabla}\times\vec{f}^{(n)}\right)^2
-\frac{1}{4}\frac{1}{4n+3}
\left((2n+1)\vec{\nabla}\times\vec{f}^{(n)}
+(2n+2)\vec{\nabla}\times\vec{f}^{(n+1)}\right)^2
\nonumber\\&&
+\frac{4n+1}{2}\left(\Pi^{(n)}\right)^2
+\frac{1}{2}\frac{1}{4n+3}
\left((2n+1)\vec{\nabla}\theta^{(n)}
+(2n+2)\vec{\nabla}\theta^{(n+1)}\right)^2
\nonumber\\&&
\left.
+\frac{m_D}{3}\vec{\nabla}\cdot\vec{A}
\left(\theta^{(0)}+2\theta^{(1)}\right)
-\frac{m_D}{3}\vec{\nabla}\times\vec{A}
\left(\vec{f}^{(0)}-\vec{f}^{(1)}\right)
+\frac{1}{6}m_D^2\vec{A}^2\right]
\Biggr\},
\labx{equ:ham}
\end{eqnarray}

\section{Lattice discretization}
\label{app:latt}

\subsection{Classical theory}
\label{app:lattcl}
In order to carry out numerical simulations described in
Section~\ref{sect:classsim}, we discretize the Hamiltonian
(\ref{equ:clHam}) and the equations of motion (\ref{equ:fulleom})
in the standard leap-frog fashion. 
The Higgs field $\phi$ was defined at the lattice
sites, and its canonical momentum at temporal links between time
slices. The gauge field $\vec{A}$ was represented by real numbers
defined at links between lattice sites, and the electric field
$\vec{E}$ by temporal plaquettes that connect these links.
Therefore $\pi_{(t,\vec{x})}$ is actually defined at the point
$(t+\delta t/2,\vec{x})$,
$A_{i,(t,\vec{x})}$ at $(t,\vec{x}+\hi/2)$ and 
$E_{i,(t,\vec{x})}$ at $(t+\delta t/2,\vec{x}+\hi/2)$.
Here $\hi$ is a vector of length $\delta x$ in the $i$ direction.

The lattice version of the Hamiltonian (\ref{equ:clHam}) is
\begin{eqnarray}
\labx{equ:lattHamcl}
H&=&\sum_{\vec x}\delta x^3\left[
\frac{1}{2}\sum_iE_{i}
+\frac{1}{2}\sum_{i}\left(\epsilon_{ijk}\Delta_j^+A_k\right)^2
+\pi^*\pi
\right.\nonumber\\&&\left.
-\frac{2}{\delta x^2}\sum_i{\rm Re}\phi_{(\vec x)}^*
U_{i,(\vec x)}\phi_{(\vec x+\hi)}
+\left(m_{T}^2+\frac{6}{\delta x^2}\right)|\phi|^2+\lambda|\phi|^4
\right].
\end{eqnarray}
where 
\begin{eqnarray}
U_{i}&=&
\exp\left(ie\delta x 
A_{i}\right),\nonumber\\
\Delta^\pm_i\phi_{(\vec{x})}&=&
\pm\delta x^{-1}\left(\phi_{(\vec{x}\pm\hi)}-\phi_{(\vec{x})}\right).
\end{eqnarray}
We also define the lattice version of the covariant derivative
\begin{eqnarray}
D^+_i\phi_{(\vec{x})}&=&\delta x^{-1}\left(
U_{i,(\vec{x})}\phi_{(\vec{x}+\hi)}
-\phi_{(\vec{x})}\right),\nonumber\\
D^-_i\phi_{(\vec{x})}&=&\delta x^{-1}\left(
\phi_{(\vec{x})}-
U^*_{i,(\vec{x}-\hi)}\phi_{(\vec{x}-\hi)}\right)
.
\end{eqnarray}

The value of the bare lattice mass $m^2_{T}$ is given by
Eq.~(\ref{equ:mTren}) and was chosen in such
a way that the Hamiltonian (\ref{equ:lattHamcl}) describes the
thermodynamics of the finite-temperature theory with renormalized 
mass $m^2$
correctly~\cite{Laine:1995ag,Laine:1998dy}.

The discretized equations of motion are
\begin{eqnarray}
\Delta_t E_{i,(t,\x)}&=&
\epsilon_{ijk}\epsilon_{klm}
\Delta^-_j\Delta^+_lA_{i,(t,\vec{x})}
-2e{\rm Im}\phi^*_{(t,\x)}D^+_i\phi_{(t,\x)},
\nonumber\\
\Delta_t\pi_{(t,\x)}&=&
D_i^-D_i^+\phi_{(t,\x)}
-m_T^2\phi_{(t,\vec{x})}
-2\lambda|\phi_{(t,\vec{x})}|^2\phi_{(t,\vec{x})},
\nonumber\\
\Delta_t A_{i,(t+\delta t,\x)}&=&-
E_{i,(t,\x)},
\nonumber\\
\Delta\phi_{(t+\delta t,\x)}&=&\pi_{(t,\x)},
\end{eqnarray}
where $\Delta_t\phi_{(t)}
=\delta t^{-1}[\phi_{(t)}-\phi_{(t-\delta t)}]$ etc.

The lattice version of the Gauss law (\ref{equ:gauss}) is
\begin{equation}
\sum_i\Delta^-_iE_{i,(t,\x)}=
2e{\rm Im}\phi^*_{(t,\vec{x})}
\pi_{(t,\vec{x})}.
\end{equation}
This is an extra constraint the initial field configuration must
satisfy.

\subsection{HTL theory}
\label{app:latthtl}
In the HTL simulations described in Section~\ref{sect:htlsim},
the soft modes were discretized in the same way as in the classical
case. 
The extra field $\theta$ is defined at lattice sites and $f_i$ 
at the plaquettes. We denote by $f_{i,(t,\vec{x})}$ the field value at 
$(t,\vec{x}+\hat{x}/2+\hat{y}/2+\hat{z}/2-\hi/2)$.

The lattice version of the HTL-improved Hamiltonian (\ref{equ:ham}) is
\begin{equation}
H_{\rm HTL}=H+\delta x^3\sum_{\x}\left[
{\cal H}_F+{\cal H}_f+{\cal H}_\Pi+{\cal H}_\theta\right],
\end{equation}\
where
\begin{eqnarray}
{\cal H}_F&=&\sum_{n=0}^\infty\frac{1}{4}\left[
\frac{1}{4n+3}
\left((2n+1)\vec{F}^{(n)}+(2n+2)\vec{F}^{(n+1)}\right)^2\right.
\nonumber\\&&\left.
-(4n+1)
\left(C^+_n\vec{F}^{(n+1)}+C^0_n\vec{F}^{(n)}
+C^-_n\vec{F}^{(n-1)}\right)^2\right],
\nonumber\\
{\cal H}_f&=&\sum_{n=0}^\infty
\left[
\frac{4n+1}{4}\left(\epsilon_{ijk}\Delta^-_jf_k^{(n)}\right)^2
-\frac{1}{4}\frac{1}{4n+3}
\left((2n+1)\epsilon_{ijk}\Delta^-_jf_k^{(n)}
\right.\right.\nonumber\\&&\left.\left.
+(2n+2)\epsilon_{ijk}\Delta^-_jf_k^{(n+1)}\right)^2\right]
-\frac{m_D}{3}\epsilon_{ijk}\Delta_j^+A_{k}
\left(f_k^{(0)}-f_k^{(1)}\right)
+\frac{1}{6}m_D^2A_i^2,
\nonumber\\
{\cal H}_\Pi&=&\sum_{n=0}^\infty
\frac{4n+1}{2}\left(\Pi^{(n)}\right)^2,
\nonumber\\
{\cal H}_\theta&=&\sum_{n=0}^\infty
\frac{1}{2}\frac{1}{4n+3}
\left((2n+1)\Delta^+_i\theta^{(n)}
+(2n+2)\Delta^+_i\theta^{(n+1)}\right)^2
+\frac{m_D}{3}\Delta^-_iA_i
\left(\theta^{(0)}+2\theta^{(1)}\right).
\labx{equ:hamlat}
\end{eqnarray}

The Gauss law (\ref{equ:gaussHTL}) can be written in the form
\begin{equation}
\Pi^{(0)}_{(t,\x)}=\frac{1}{m_D}\left(
\sum_i\Delta^-_iE_{i,(t,\x)}
-2e{\rm Im}\phi^*_{(t,\vec{x})}\pi_{(t,\vec{x})}
\right),
\end{equation}
and therefore we can eliminate $\Pi^{(0)}$ from the Hamiltonian.

The bare Higgs mass has the value given in Eq.~(\ref{equ:mTren}),
and the Debye mass is~\cite{Laine:1995ag,Laine:1998dy}
\begin{equation}
m_D^2=\frac{1}{3}e^2T^2-2e^2\frac{3.176 T}{4\pi\delta x}
.
\end{equation}
These counterterms were calculated by matching static correlators and
in the absence of Lorenz invariance, they do not remove all the
ultraviolet divergences from real-time quantities. However, since
our lattice spacing is relatively large, this leads only to small errors.

The discretized equations of motion are
\begin{eqnarray}
\Delta_t E_{i,(t,\x)}&=&
\epsilon_{ijk}\epsilon_{klm}
\Delta^-_j\Delta^+_lA_{i,(t,\vec{x})}
-2e{\rm Im}\phi^*_{(t,\x)}D^+\phi_{i,(t,\x)}
\nonumber\\&&
-\frac{m_D}{3}\left(
\Delta^+_i\theta^{(0)}_{(t,\x)}
+2\Delta^+_i\theta^{(1)}_{(t,\x)}
+\epsilon_{ijk}\Delta^-_jf^{(0)}_{k,(t,\x)}
-\epsilon_{ijk}\Delta^-_jf^{(1)}_{k,(t,\x)}
\right),
\nonumber\\
\Delta_t\pi_{(t,\x)}&=&
D_i^-D_i^+\phi_{(t,\x)}
-m_T^2\phi_{(t,\vec{x})}
-2\lambda|\phi_{(t,\vec{x})}|^2\phi_{(t,\vec{x})},
\nonumber\\
\Delta_t F_{i,(t,\x)}^{(n)}&=&
-\epsilon_{ijk}\epsilon_{klm}\Delta^+_j\Delta^-_lf_{m,(t,\x)}^{(n)}
+\delta_{n,0}m_D\epsilon_{ijk}\Delta^+_jA_{k,(t,\x)},
\nonumber\\
\Delta_t \Pi_{(t,\x)}^{(n)}&=&
C^+_n\Delta_i^+\Delta_i^-\theta^{(n+1)}_{(t,\x)}
+C^0_n\Delta_i^+\Delta_i^-\theta^{(n)}_{(t,\x)}
+C^-_n\Delta_i^+\Delta_i^-\theta^{(n-1)}_{(t,\x)}
\nonumber\\&&
-m_D\Delta_i^-A_{i,(t,\x)}
\left(\frac{1}{3}\delta_{n,0}+\frac{2}{15}\delta_{n,1}
\right),
\nonumber\\
\Delta_t A_{i,(t+\delta t,\x)}&=&-
E_{i,(t,\vec{x})},
\nonumber\\
\Delta_t\phi_{(t+\delta t,\x)}&=&\pi_{(t,\x)},
\nonumber\\
\Delta_t f_{i,(t+\delta t,\x)}^{(n)}
&=&
C^{+}_nF_{i,(t,\x)}^{(n+1)}
+C^{0}_nF_{i,(t,\x)}^{(n)}
+C^{-}_nF_{i,(t,\x)}^{(n-1)},
\nonumber\\
\Delta_t\theta_{(t+\delta t,\x)}^{(n)}&=&\Pi_{(t,\x)}^{(n)}.
\end{eqnarray}

\end{document}